# Clocking Auger electrons

D. C. Haynes[1,2,3], M. Wurzer[4], A. Schletter[4], A. Al-Haddad[5,6], C. Blaga[7,8], C. Bostedt[5,6,9], J. Bozek[10], M. Bucher[6], A. Camper[7], S. Carron[11], R. Coffee[11], J. T. Costello[12], L. F. DiMauro[7], Y. Ding[11], K. Ferguson[5], I. Grguraš[1,2], W. Helml[4,13], M. C. Hoffmann[11], M. Ilchen[14,15], S. Jalas[16], N. M. Kabachnik[14,17], A. K. Kazansky[18], R. Kienberger[4], A. R. Maier[2,16], T. Maxwell[5], T. Mazza[14], M. Meyer[14], H. Park[7], J. S. Robinson[11], C. Roedig[7], H. Schlarb[20], R. Singla[1,2], F. Tellkamp[1,2], K. Zhang[7], G. Doumy[19], C. Behrens[20], A. L. Cavalieri[1,2,3,5,21]

---

[1] *Max Planck Institute for the Structure and Dynamics of Matter, Luruper Chausee 149, 22761 Hamburg, Germany*
[2] *Center for Free-Electron Laser Science, Luruper Chaussee 149, 22761 Hamburg, Germany*
[3] *The Hamburg Centre for Ultrafast Imaging, Universität Hamburg, Mittelweg 177, 20148 Hamburg, Germany*
[4] *Physics Department, Technische Universität München, James Franck Straße, 85748 Garching, Germany*
[5] *Paul Scherrer Institute, Forschungsstrasse 111, 5232 Villigen, Switzerland*
[6] *Argonne National Laboratory, 9700 Cass Avenue, Lemont, IL 60439, USA*
[7] *The Ohio State University, Columbus, OH 43210, USA*
[8] *Kansas State University, Manhattan, KS 66506, USA*
[9] *LUXS Laboratory for Ultrafast X-ray Sciences, Institute of Chemical Sciences and Engineering, Ecole Polytechnique Fédérale de Lausanne (EPFL), CH-1015 Lausanne, Switzerland*
[10] *Synchrotron SOLEIL, l'Orme des Merisiers, Saint-Aubin, BP 48, 91192 Gif-sur-Yvette, France*
[11] *Linac Coherent Light Source/SLAC National Accelerator Laboratory, 2575 Sand Hill Rd, Menlo Park, CA 94025, USA*
[12] *National Center for Plasma Science and Technology and School of Physical Sciences, Dublin City University, Dublin, Ireland*
[13] *Technische Universität Dortmund, Maria-Goeppert-Mayer-Straße 2, 44227 Dortmund, Germany*
[14] *European XFEL GmbH, Holzkoppel 4, 22869 Schenefeld, Germany*
[15] *Institut für Physik und CINSaT, Universität Kassel, Heinrich-Plett-Str. 40, 34132 Kassel, Germany*
[16] *Department of Physics, Universität Hamburg, Luruper Chaussee 149, 22761 Hamburg, Germany*
[17] *Skobeltsyn Institute of Nuclear Physics, Lomonosov Moscow State University, Moscow 119991, Russia*
[18] *Departamento de Fisica de Materiales, University of the Basque Country UPV/EHU; Donostia International Physics Center (DIPC), E-20019 San Sebastian/Donostia, Basque Country; IKERBASQUE, Basque Foundation for Science, E-48011 Bilbao, Spain*
[19] *Chemical Sciences and Engineering Division, Argonne National Laboratory, 9700 S. Cass Avenue, Lemont, IL, 60439, USA*
[20] *Deutsches Elektronen-Synchrotron, Notkestraße 85, 22607 Hamburg, Germany*
[21] *Institute of Applied Physics, University of Bern, Sidlerstr. 5, 3012 Bern, Switzerland*

**Intense X-ray free-electron lasers (XFELs) can rapidly excite matter, leaving it in inherently unstable states that decay on femtosecond timescales. As the relaxation occurs primarily via Auger emission, excited state observations are constrained by Auger decay.** *In situ* **measurement of this process is therefore crucial, yet it has thus far remained elusive at XFELs due to inherent timing and phase jitter, which can be orders of magnitude larger than the timescale of Auger decay. Here, we develop a new approach termed** *self-referenced attosecond streaking,* **based upon simultaneous measurements of streaked photo- and Auger electrons. Our technique enables sub-femtosecond resolution in spite of jitter. We exploit this method to make the first XFEL time-domain measurement of the Auger decay lifetime in atomic neon, and, by using a fully quantum-mechanical description, retrieve a lifetime of $2.2^{+0.2}_{-0.3}$ fs for the KLL decay channel. Importantly, our technique can be generalised to permit the extension of attosecond time-resolved experiments to all current and future FEL facilities.**

The motion of electrons underpins many of the fastest processes in atomic, molecular, and condensed matter systems. In recent decades, electron transport has been the subject of intense scrutiny, thanks in large part to concurrent advances in ultrafast lasers and corresponding spectroscopic techniques. Still more recently, the exploitation of high-harmonic generation (HHG)-based extreme ultraviolet (XUV) sources has enabled the interrogation of matter with unprecedented time resolution. However, XUV sources lack the requisite intensity to create highly excited states of matter, many of which are driven by multi-photon processes. The advent of XFELs – which occurred in parallel to advances in table-top XUV sources – has now made it feasible to excite and investigate these states, leading to pioneering techniques including double-core-hole spectroscopy[1] and the XFEL-pumped X-ray laser[2,3]. In many cases the evolution of such highly excited systems can be characterised in terms of short-lived core holes and their decay.

Auger decay is a fundamental manifestation of correlated electron dynamics, wherein the action of one electron affects another. In this process a tightly bound electron in an atom or molecule is ejected, either by absorption of an X-ray photon or collision with an energetic particle. When a more weakly bound electron fills the resulting core-hole, the energy released by this relaxation process can induce ejection of another electron, known as an Auger electron[4].

This non-radiative process is the dominant decay mechanism for elements with a low atomic number. In these cases, the core-hole decay lifetime is essentially equivalent to the Auger decay lifetime, and on the order of femtoseconds[5-7]. When the core-hole is created by photoionisation, the Auger decay lifetime is related by the uncertainty principle to the spectral line width of the photoemission line[7-9]. High-resolution electron spectra, mostly measured using high-brightness synchrotron sources[4,8,10], have therefore been used to infer Auger decay lifetimes.

Alternatively, it is possible to access these dynamics directly in the time domain, using X-ray pulses with a duration comparable to, or shorter than, the Auger decay lifetime. For example, in proof-of-principle experiments in krypton[5] using relatively weak, table-top attosecond XUV pulses[11-13], core-holes were created impulsively by photoionisation. The subsequent Auger decay was then temporally resolved by dressing the electron emission with an optical laser pulse. In these experiments, the photoemission profile matches the temporal profile of the exciting attosecond XUV pulse, as photoemission occurs on even shorter attosecond timescales[14]. In contrast, the Auger emission occurs over a longer duration – typically on a femtosecond timescale.

In a number of attosecond investigations of Auger decay[15-18], the temporal profile of Auger emission has been approximated by a convolution of the XUV pulse profile with an exponential decay curve. This phenomonological *ad hoc* model is based on a two-step description of Auger decay, treating the ionisation and subsequent Auger emission as distinct processes. More recently an alternative model has been proposed, in which the process is treated with a consistent, fully quantum-mechanical description[19-20]. The newer model treats Auger decay in terms of the amplitudes of the states involved, rather than in terms of those states' populations. In the case where the exciting pulse duration is comparable to or longer than the Auger decay lifetime, a substantial difference appears in the emission profiles predicted by the two models.

In this paper, we present a new experimental study of Auger decay emission in the time domain, using intense, femtosecond soft X-ray pulses from an XFEL of duration commensurate with the core-hole lifetime. We find that only the quantum-mechanical treatment can produce satisfactory agreement with our data, thus highlighting the limits of the *ad hoc* model. This represents the first experimental demonstration of the effect of quantum coherence of photo- and Auger electrons.

**Ultrafast science at XFELs**

XFELs provide extremely intense pulses across the soft and hard X-ray spectral domains, permitting the interrogation of a wide range of systems that are not accessible via other X-ray light sources. In principle, the pulses delivered at XFELs can be short enough to explore few- or even sub-femtosecond dynamics[21], including Auger decay[22]. Furthermore, XFEL pulses are many orders of magnitude more intense than X-rays from other sources, and consequently can be used to pump and probe highly excited states of matter[1-3,23-29], many of which are constrained or influenced by Auger decay.

Despite these favourable characteristics, existing attosecond time-resolved spectroscopies have hitherto been impossible to apply at XFELs. Even with modern electronic and optically distributed reference signals[30], it is currently unfeasible to perfectly synchronise an XFEL pulse with the field of an external streaking laser pulse; experiments at XFELs suffer from ever-present timing and phase jitter which place limits on the achievable time resolution. Independent time-of-arrival measurements can be used in post-processing to dramatically improve the effective time resolution[31-36], but their implementation is challenging, and in many cases the jitter remains more than an order of magnitude larger than the timescales of Auger decay. Recent developments of angular streaking at XFELs have shown promise as a diagnostic tool for mitigating jitter[37], but the complexity of these techniques has thus far precluded their broader application for experimental measurements. As a result, direct time-resolved studies of most electron dynamics at XFELs have generally not yet been accomplished. There is thus a need for a straightforward technique which can unite the advantages of two very disparate light sources: whilst XFELs are the only sources able to deliver intense, ultrashort X-ray pulses and create highly excited states of matter, it is so far primarily table-top attosecond sources that have been able to provide adequate time resolution to probe the electronic dynamics underpinning those states.

Here, we develop and utilise a new self-referenced streaking approach that circumvents timing jitter and allows for the extension of table-top attosecond spectroscopy to XFELs. This will facilitate a new class of experiments benefitting from highly intense X-ray pulses alongside attosecond time resolution. As a first demonstration, we have measured the KLL Auger decay lifetime in atomic neon, in the time domain and with sub-femtosecond precision, paving the way for the extension of the technique to a variety of ultrafast measurements at XFELs worldwide.

**Time-resolved electron spectroscopy**

A schematic of the experimental apparatus at the Linac Coherent Light Source (LCLS) FEL is shown in Figure 1. We induce 1s core-level photoemission and subsequent Auger decay with an XFEL pulse whose photon energy is centred at 1130 eV. The pulse is directed into a dilute neon gas target, and the photo- and Auger electrons are analysed using a time-of-flight spectrometer equipped with an electrostatic lens to increase the collection efficiency.

We perform single-shot measurements of both peaks simultaneously, taking advantage of the fact that the KLL Auger spectrum of neon is dominated by the strong emission line associated with the $Ne^{2+}$ $2p^4$ $^1D_2$ final state[38].

An important prerequisite for streaking measurements is that the dynamics must occur within a half-cycle of the streaking laser field[11,39]. Based on the peak current in the bunch compressor, the X-ray pulse duration in our experiment was estimated to be under 10 fs full-width at half-maximum (FWHM), so that an infrared (IR) streaking field with a correspondingly long optical cycle is required. To this end, a Titanium-Sapphire 800 nm femtosecond laser is used as the pump source to generate IR signal and idler pulses in a barium borate crystal via optical parametric amplification. These pulses are mixed in a gallium selenide crystal for difference-frequency generation, producing mid-infrared (MIR) streaking pulses with a wavelength of 17 µm. The streaking period, therefore, is 57 fs, so we can be confident that the exciting X-ray pulse and the few-femtosecond Auger dynamics will be fully encompassed within a 24 fs half-cycle of the streaking field.

The linearly polarised streaking laser has a time-dependent electric field

$$E_{IR}(t) = E_0(t) \cos(\omega_{IR} t), \quad (1)$$

and vector potential

$$A(t) = \frac{-E_0(t)}{\omega_{IR}} \sin(\omega_{IR} t), \quad (2)$$

such that $E_{IR}(t) = -\frac{\partial A}{\partial t}$. The symbols $E_0(t)$ and $\omega_{IR}$ represent the slowly varying amplitude of the streaking field and its angular frequency respectively.

Upon interaction with the streaking laser field, the emitted electrons' change in kinetic energy $\Delta E$ can be approximated by

$$\Delta E \approx \sin(\phi_i) \sqrt{8 E_{el} U_p}, \quad (3)$$

where $\phi_i$ is the phase of the streaking pulse at the moment of photoemission $t_i$ and $E_{el}$ is the electrons' field-free kinetic energy[40,41]. The ponderomotive potential $U_p$ is given by

$$U_p = \frac{e^2 E_0^2}{4 m_e \omega_{IR}^2}, \tag{4}$$

where $e$ and $m_e$ are the charge and mass of the electron. Examination of equations (2), (3) and (4) reveals that the change in the final change in kinetic energy $\Delta E$ experienced by the observed electron is proportional to the vector potential $A(t)$ of the streaking field at the moment of interaction. For extended emission the streaking laser in effect maps the time domain onto the sheared electron spectrum. This approach provides a route to reconstruct the temporal characteristics of the electron emission with the potential for attosecond resolution[11,40]. A crucial requirement is that the amplitude and phase of the streaking field acting on the emitted electrons must be known with sufficient precision.

**Self-referenced streaking spectroscopy**

It is possible to determine both the amplitude and phase of the streaking laser field for each shot solely by observing shifts in the kinetic energy spectrum, but at least two distinct measurements are required. In this experiment, we observe the streaked energies of both the photoemission and Auger peaks. Note that, whilst the photo- and Auger electrons are both shifted in kinetic energy according to equation (3), the Auger electrons are generally emitted later than the photoelectrons. As a result, the phase of the streaking laser will have advanced by some amount in the time between emissions due to the pulse's propagation through the stationary target. Consequently, the Auger electrons' energy shift will be a function not only of $\emptyset_i$, but of $\emptyset_i + \emptyset_A$, where $\emptyset_A$ represents the phase advance between the instants of photo- and Auger emission. The result is that the change in energy experienced by photoelectrons and Auger electrons will generally differ in magnitude and even sign, depending on the temporal overlap and absolute carrier-envelope phase (CEP) of the MIR pulse[I].

Conceptually, if one were to smoothly vary the overlap between X-ray and CEP-stable streaking pulses, the sinusoidal curves traced out by the two emission peaks' centres of energy would be temporally displaced by the time elapsed between the two events, as illustrated in Figure 2a and Figure 2b. In effect, each peak in the electron energy spectrum independently samples the oscillation of the streaking vector potential. When plotting the two streaked centres of energy against each other, as in Figure 2c, the resultant ring has an ellipticity determined by the phase shift between the two sine curves. A phase shift of 0 (i.e. the case where both emissions were simultaneous) would result in a straight line, as both emissions would experience the same vector potential in each shot. If the shift was $\frac{\pi}{2}$, the graph would be a wide ellipse with major and minor axes parallel to those of the coordinate system, because whenever one emission interacted with a zero crossing of the streaking field, the other would interact with an extremum. Finally, a phase shift between 0 and $\frac{\pi}{2}$ would lead to a sheared ellipse.

In reality, the CEP of the MIR streaking pulse cannot be controlled during experiments at XFELs. Therefore, each single-shot measurement is made with a random streaking phase. Nevertheless, if a large enough set of measurements is accumulated, the entire parameter space will be explored, and a scatter plot forming an ellipse like that in Figure 2c can be constructed.

In addition to its randomly varying phase, the precise arrival time of the streaking pulse with respect to the X-ray pulse fluctuates, resulting in a normal distribution of arrival times. Therefore, the strength of the streaking effect varies from shot to shot, depending on the temporal overlap between the X-ray pulse and MIR intensity pulse envelope. Whilst variation in the streaking field phase leads to the characteristic ellipse, timing-jitter-induced variations in streaking strength result in a broadening of the elliptical distribution, since for any given angle around the ellipse there are a range of possible displacements from its centre.

---

[I] It should be noted that since the field-free kinetic energies $E_{el}$ of the two types of electrons are different, there would always be a difference in the magnitude of their energy changes, even if the emissions were simultaneous. This is due to the factor of $\sqrt{E_{el}}$ in equation (3).

The ellipse makes it simple to identify those shots for which the photoemission burst coincides with a zero crossing of the streaking vector potential. Such shots appear on the 'equator' of the ellipse, since the photoelectrons experienced a minimal energy shift. After identifying these shots, it is possible to calculate the duration of the X-ray pulse by comparing the width of the photoemission line in these maximally broadened shots to that measured in the absence of laser field. Using this method, we determined the average X-ray pulse FWHM to be $7 \pm 1$ fs, as described in the Supplementary Information.

The elliptical distribution, generated by correlating the streaking-induced shift in kinetic energy of the photoelectron and Auger peaks in single-shot measurements made over a complete set of streaking field parameters, is the key to our technique. In effect, the correlation plot is a map, with each position pinned to a unique set of streaking field parameters. Once this distribution has been obtained, all subsequent single-shot measurements can be mapped to retrieve the instantaneous streaking phase and amplitude. Single-shot measurements performed with desired streaking parameters can be identified and isolated, even though those parameters are uncontrolled during the experiment. This is how we have extended the techniques of table-top attosecond spectroscopy to be applied at XFELs, granting a dramatic increase in achievable time resolution which is ultimately limited only by the X-ray pulse duration.

**The Auger decay lifetime in neon**

The streaked kinetic energies of the photoelectron and Auger electron peaks are determined by numerically fitting the recorded spectra in each single-shot measurement and calculating the centre of energy of each peak. By comparing the streaked energies to the corresponding field-free values, we determine the changes in kinetic energy, $\Delta E_{1s}$ and $\Delta E_{Auger}$, induced by the streaking field in each single-shot measurement. As shown in Figure 3, after making many thousands of measurements, correlating $\Delta E_{1s}$ against $\Delta E_{Auger}$ for every pair of single-shot measurements reveals an elliptical distribution. Specific regions of the ellipse are highlighted, with sketches of the corresponding measurement conditions shown in the subplots on the right. The subplots indicate how the correlation map can be used to navigate to previously inaccessible streaking parameters: the angular coordinate of each point identifies the streaking phase for that shot, and its radial coordinate is a function of the streaking field amplitude.

As an alternative to examining features in the individual or averaged streaked Auger spectra, which is not possible here due to limited energy resolution, the degree of ellipticity in the distribution can provide access to the Auger decay lifetime. Note that any ellipse can be described using the set of parametric equations

$$x(\theta) = A \sin(\theta + \phi_A), \qquad y(\theta) = B \sin(\theta). \tag{5}$$

In our case, $x$ and $y$ correspond to the change in kinetic energy of the Auger and photoelectrons respectively. As described in detail in the Supplementary Information, the angle $\phi_A$ is the phase advance that occurs between the two instants of electron emission. It is given by

$$\phi_A = \sin^{-1}\left(\frac{y_1}{y_2}\right), \tag{6}$$

where $y_1$ is the ellipse's $y$-intercept, and $y_2$ is its maximum value of $y$. By examining the angular sectors of the ellipse corresponding to $y_1$ and $y_2$ we can calculate them, obtaining values of $8.0 \pm 0.1$ eV and $20.9 \pm 0.1$ eV respectively. Using these values in conjunction with equation (6) enables us to calculate the phase advance $\phi_A$ to be $0.39 \pm 0.01$ radians.

Details of the selection of the sectors containing the points, and the calculation of the uncertainty on these values, can be found in the Supplementary Information. The corresponding time-delay, $\tau_{delay}$, between the photo and Auger emission bursts can then be calculated using the observed phase shift $\phi_A$ and $T_{IR} = 56 ^{+3}_{-7}$ fs, the period of the streaking pulse:

$$\tau_{delay} = \frac{\phi_A}{2\pi} T_{IR}. \tag{7}$$

Applying this algorithm to the distribution shown in Figure 3, we obtain a delay of $3.5 ^{+0.3}_{-0.5}$ fs. The sub-femtosecond uncertainty on this value was obtained by propagating the uncertainties on $\phi_A$ and $T_{IR}$ using a standard functional approach, as detailed further in the Supplementary Information. If we were to assume an *ad hoc* two-step description of Auger decay, as has been done in the past[5], this delay would correspond exactly to the Auger decay lifetime: This model, as discussed in the Supplementary Information, predicts that the delay between the weighted centres of the photo and Auger temporal emission profiles is identical to its decay lifetime.

However, under our experimental conditions, where the X-ray pulse duration is comparable or longer than the expected Auger decay lifetime, the *ad hoc* model is expected to break down[19], and we must turn to a more precise quantum-mechanical description[20] of Auger decay in order to correctly interpret our results. This comprehensive model of our experiment shows that the observed delay between the two electron emission bursts is markedly larger than the Auger decay lifetime. The full model detailed in the Supplementary Information allows us to assume a variety of possible Auger decay lifetimes and calculate the corresponding delay $\tau_{delay}$ that would be observed in each case. Incorporating the X-ray pulse duration and streaking wavelength used in our experiments reveals a notable difference between $\tau_{delay}$ and the assumed Auger decay lifetime, but the model allows us to map between the two quantities. This procedure led to the conclusion that the observed delay $\tau_{delay} = 3.5 ^{+0.3}_{-0.5}$ fs between the emission bursts corresponds to a true Auger decay lifetime of $2.2 ^{+0.2}_{-0.3}$ fs. This is in agreement with measurements reported from spectral linewidth studies, which have found values in the range 2.0-2.6 fs[7,10,43,44].

The significant discrepancy between $\tau_{delay}$ and the Auger decay lifetime – which are assumed to be identical in the *ad hoc* model – demonstrates the necessity for a full quantum-mechanical treatment for experiments such as ours, where the exciting X-ray pulse duration is comparable to or longer than the Auger decay lifetime.

**Conclusion and outlook**

This measurement, the first of its type to be performed at an XFEL, was made possible via self-referenced attosecond streaking, a novel experimental technique. Following this successful demonstration of its efficacy, self-referenced streaking will enable experimentalists to take advantage of the extreme-intensity X-ray pulses at XFELs while simultaneously exploiting the unrivalled time resolution provided by attosecond streaking spectroscopy.

In conjunction with the new technique, the measurement was made possible via the application of a consistent quantum model of Auger decay[20]. Through the application of this more advanced model, we demonstrated that the older *ad hoc* model significantly overestimates the extracted lifetime under the present experimental conditions. This will have major ramifications for future studies of Auger decay, especially those applying our new experimental techniques to make the measurement at XFELs.

Because so many highly excited states of matter relax via Auger decay, this result may also help to inform future studies on double-core-hole spectroscopy, XFEL-pumped X-ray lasers, and other innovative techniques dependent upon the timescales of Auger processes. Beyond simple atomic systems, our self-referenced Auger measurements could pave the way for investigations into the effect of a system's chemical environment on Auger decay[45]; a comparison could, for example, be made between decay rates of carbon in $CF_4$ and $CO$[6]. Furthermore, we expect that precise temporal characterisation of Auger decay processes in complex systems will be crucial in interpreting diffraction and scattering patterns in single-molecule imaging experiments, where a significant proportion of Auger electrons are known to deposit energy into molecular samples after emission[46,47].

**References:**

1. Young, L. *et al*. Femtosecond electronic response of atoms to ultra-intense X-rays. *Nature* **466**, 56, (2010).
2. Rohringer, N. An atomic inner-shell laser pumped with an X-ray free-electron laser. *J. Phys.: Conf. Ser.* **194** 012012 (2009).
3. Rohringer, N. *et al.* Atomic inner-shell X-ray laser at 1.46 nanometres pumped by an X-ray free-electron laser. *Nature* **481**, 488-491 (2012).
4. Mehlhorn, M. 70 years of Auger spectroscopy, a historical perspective. *Journal of Electron Spectroscopy and Related Phenomena* **93**,1 (1998).
5. Drescher, M. *et al.* Time-resolved atomic inner-shell spectroscopy. *Nature* **419**, 803 (2002).
6. Carroll, T. *et al.* Carbon 1s photoelectron spectroscopy of CF4 and CO: Search for chemical effects on the carbon 1s hole-state lifetime. *J. Chem. Phys.* **116**, 10221 (2002).
7. Coreno, M., *et al*. Measurement and ab initio calculation of the Ne photoabsorption spectrum in the region of the K edge. Phys. Rev. A **59**, 2494 (1999)
8. Jurvansuu, M. *et al.* Inherent lifetime widths of Ar $2p^{-1}$, Kr $3d^{-1}$, Xe $3d^{-1}$, and Xe $4d^{-1}$ states. *Phys. Rev. A* **64**, 012502 (2001).
9. Krause, M. O. Atomic radiative and radiationless yields for *K* and *L* shells. *J. Phys. Chem. Ref. Data* **8,** 307 (1979).
10. Schmidt, V. *et al. Electron Spectrometry of Atoms Using Synchrotron Radiation*. (Cambridge University press, Cambridaage, 1997).
11. Kienberger, R. *et al.* Atomic transient recorder. *Nature* **427**,817–821 (2004).
12. Goulielmakis, E. *et al.* Single-cycle nonlinear optics. *Science* **320**,1614–1617 (2008).
13. Hentschel, M., *et al.* Attosecond metrology. *Nature* **414**, 509-513 (2001).
14. Schultze, M. *et al.* Delay in Photoemission. *Science* **328**, 1658-1662 (2010).
15. Uiberacker, M. *et al*. Attosecond real-time observationof electron tunnelling in atoms. *Nature* **446**, 627 (2007).
16. Uphues, T., *et al*. Ion-charge-state chronoscopy of cascaded atomic Auger decay. New J. Phys. **10**, 25009 (2008).
17. Krikunova, M. *et al.* Time-resolved ion spectrometry on xenon with the jitter-compensated soft x-ray pulses of a free-electron laser. *New J. Phys.* **11**, 123019 (2009).
18. Verhoef, A. J. *et al.* Time- and energy-resolved measurement of Auger cascades following Kr 3d excitation by attosecond pulses. *New J. Phys.* **13**, 113003 (2011)
19. Smirnova, O., Yakovlev V. S., and Scrinzi, A. Quantum Coherence in the Time-Resolved Auger Measurement. *Phys. Rev. Lett.* **91**, 253001 (2003)
20. Kazansky, A. K., Sazhina, I. P., and Kabachnik, N. M. Time-dependent theory of Auger decay induced by ultra-short pulses in a strong laser field. *J. Phys. B: At. Mol. Opt. Phys.* **42**, 245601 (2009).
21. Duris, J. *et al.* Tunable isolated attosecond X-ray pulses with gigawatt peak power from a free-electron laser. *Nat. Photonics* **14**, 30–36 (2020).
22. Bostedt, C. *et al.* Linac Coherent Light Source: The first five years. *Rev. Mod. Phys.* **88**, 015007 (2016).
23. Yoneda, H. *et al.* Atomic inner-shell laser at 1.5-ångström wavelength pumped by an X-ray free-electron laser. *Nature* **524**, 446-449 (2015).
24. Fang, L. *et al.* Double core-hole production in $N_2$: beating the Auger clock. *PRL* **105**,083005 (2010).
25. Cederbaum, L. S. *et al.* On double vacancies in the core. *J. Chem. Phys.* **85**, 6513 (1986).
26. Santra, R., Kryzhevoi, N. V. & Cederbaum, L.S. X-ray Two-Photon Photoelectron Spectroscopy: A Theoretical Study of Inner-Shell Spectra of the Organic Para-Aminophenol Molecule. *Phys. Rev. Lett.* **103**, 013002 (2009).
27. Cederbaum, L. S. *et al.* Double vacancies in the core of benzene. *J. Chem. Phys.* **86**, 2168 (1987).
28. Tashiro, M. *et al.* Molecular double core hole electron spectroscopy for chemical analysis. *J. Chem. Phys.* **132**, 184302 (2010).
29. Salén, P. *et al.* Experimental verification of the chemical sensitivity of two-site double core-hole states formed by an X-ray free-electron laser. *Phys. Rev. Lett.* **108**, 15 (1991).
30. Schulz, S. *et al.* Femtosecond all-optical synchronization of an X-ray free-electron laser. *Nature Communications* **6**, 5938 (2015).
31. Schorb, S. *et al.* X-ray optical cross-correlator for gas-phase experiments at the Linac Coherent Light Source free-electron laser. *Appl. Phys. Lett.* **100**, 121107 (2012).

32. Grguraš, I. *et al.* Ultrafast X-ray pulse characterization at free-electron lasers. *Nature Photonics* **6**, 852–857 (2012).
33. Cavalieri, A. L. *et al.* Clocking femtosecond X-rays. *Phys. Rev. Lett.* **94**, 114801 (2005).
34. Bionta, M. R. *et al.* Spectral encoding of X-ray/optical relative delay. *Opt. Express* **19**, 21855–21865 (2011).
35. Harmand, M. *et al.* Achieving few-femtosecond time-sorting at hard X-ray free-electron lasers. *Nature Photonics* **7**, 215–218 (2013).
36. Hartmann, N. *et al.* Sub-femtosecond precision measurement of relative X-ray arrival time for free-electron lasers. *Nature Photonics* **8**, 706–709 (2014).
37. Hartmann, N. *et al.* Attosecond time–energy structure of X-ray free-electron laser pulses. *Nature Photonics* **12**, 215–220 (2018).
38. Krause, M. O., Stevie, F. A., Lewis, L. J., Carlson, T. A., and Moddeman, W. E. Multiple excitation of neon by photon and electron impact. *Phys. Lett. A* **31**, 81-82 (1970).
39. Frühling, U. *et al.* Single-shot terahertz-field-driven streak camera. *Nature Photonics* **3**, 523-528 (2009).
40. Itatini, J. *et al.* Attosecond streak camera. *Phys. Rev. Lett.* **88**, 173903 (2002).
41. Mazza, T. *et al.* Sensitivity of nonlinear photoionization to resonance substructure in collective excitation. *Nature Communications* **6**, 6799 (2015).
42. Helml, W. *et al.* Measuring the temporal structure of few-femtosecond free-electron laser X-ray pulses directly in the time domain. *Nature Photonics* **8**, 950 (2014).
43. Svensson, S. *et al.* Lifetime Broadening and CI-Resonances Observed in ESCA. *Physica Scripta* **14**, 141 (1976).
44. Southworth, S. H. *et al.* Double K-shell photoionization of neon. *Phys. Rev. A* **67**, 062712 (2003).
45. Cutler, J. N. *et al.* Chemical dependence of core-level linewidths and ligand-field splittings: High-resolution core-level photoelectron spectra of I 4d levels. *Physical Review Letters* **67**, 12 (1991).
46. Neutze, R. *et al.* Potential for biomolecular imaging with femtosecond X-ray pulses. *Nature* **406**, 752 (2000).
47. Gorobtsov, O. Yu., Lorenz, U., Kabachnik, N. M., and Vartanyants, I. A. Theoretical study of electronic damage in single-particle imaging experiments at x-ray free-electron lasers for pulse durations from 0.1 to 10 fs. *Phys. Rev. E* **91**, 062712 (2015).

**Acknowledgements:**

We would like to thank the staff at LCLS for preparing and operating the FEL. Use of the Linac Coherent Light Source, SLAC National Accelerator Laboratory, is supported by the US Department of Energy, Office of Science, Office of Basic Energy Sciences, under contract no. DE-AC02-76SF00515. A.L. Cavalieri, M. Meyer and D.C. Haynes acknowledge funding through the Clusters of Excellence 'The Hamburg Centre for Ultrafast Imaging' and 'CUI: Advanced Imaging of Matter' of the Deutsche Forschungsgemeinschaft (DFG) – EXC 1074 – project ID 194651731 and EXC 2056 – project ID 390715994 respectively. J. T. Costello acknowledges support by Science Foundation Ireland Grant No. 16/RI/3696 and SEAI Grant No. 19/RDD/556. L.F. DiMauro and C. Roedig acknowledge support from the National Science Foundation under Grant No. 1605042 and the U.S. Department of Energy under grant DE-FG02-04ER15614. W. Helml acknowledges financial support from a Marie Curie International Outgoing Fellowship and by the BaCaTeC program. M. Ilchen acknowledges funding from the Volkswagen Foundation within a Peter Paul Ewald-Fellowship. N. K. Kabachnik acknowledges hospitality and financial support from the theory group in cooperation with the SQS work package of European XFEL (Hamburg). A.R. Maier and S. Jalas acknowledge funding through BMBF grant 05K16GU2. T. Mazza and M. Meyer acknowledge support by Deutsche Forschungsgemeinschaft Grant No. SFB925/A1. G. Doumy was supported by the U.S. Department of Energy, Office of Science, Basic Energy Sciences, Chemical Sciences, Geosciences, and Biosciences Division under contract DE-AC02-06CH11357.

**Author Contributions:**

L.F. DiMauro and G. Doumy proposed measurement of Auger lifetimes at LCLS and A.L. Cavalieri conceived of the self-referenced streaking approach that enabled these measurements. The experiment and supporting experiments were performed by A. Al-Haddad, C. Bostedt, J. Bozek, M. Bucher, A. Camper, S. Carron, A.L. Cavalieri, R. Coffee, J.T. Costello, L.F. DiMauro, G. Doumy, K. Ferguson, I. Grguras, D.C. Haynes, M.C. Hoffmann, M. Ilchen, R. Kienberger, A.R. Maier, F. Tellkamp, W. Helml, T. Mazza, M. Meyer, H. Park, C. Roedig, A. Schletter, M. Wurzer and K. Zhang. The mid-infrared laser for streaking was provided by M.C. Hoffmann and J. Robinson. Electron beam tuning and diagnostics were provided by Y. Ding, T. Maxwell and C. Behrens. D.C. Haynes, C. Bostedt,

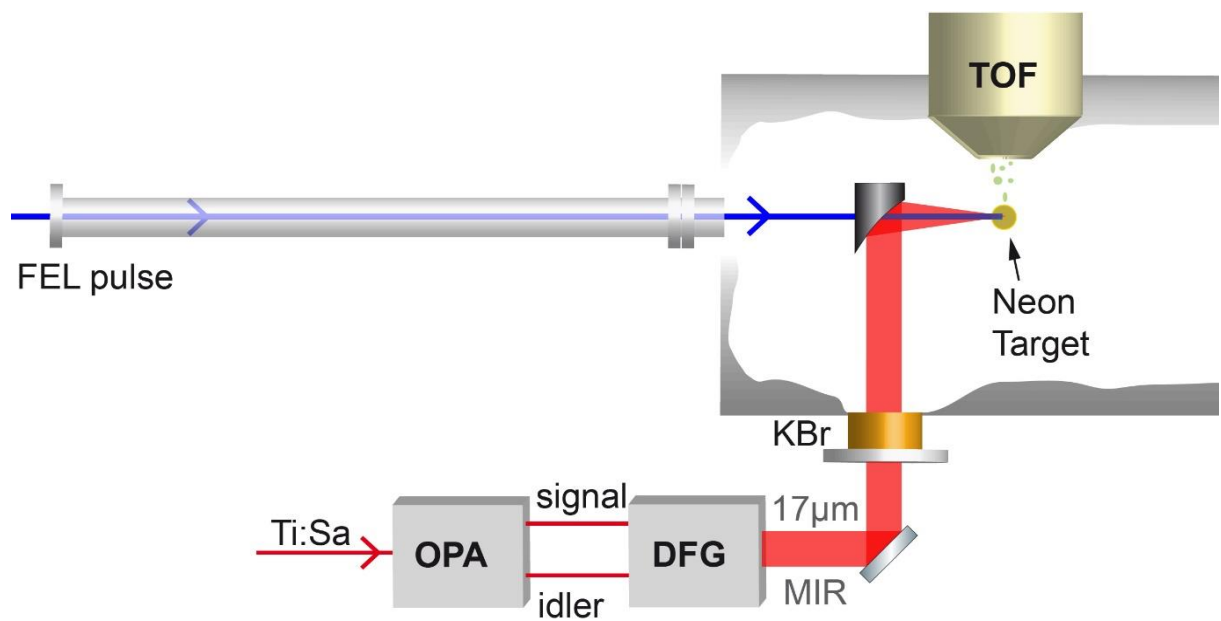

**Figure 1| Mid-Infrared Streaking.** 17 µm mid-IR (MIR) streaking laser pulses are generated by downconversion of a near-IR Titanium:Sapphire laser pulse, using an optical parametric amplifier (OPA) and difference frequency generation (DFG), and coupled into a chamber through a potassium bromide (KBr) window. The MIR pulses are focused with a 100 mm focal length parabola and overlapped with 9.2 fs 1130eV FEL pulses in a neon gas target. The resultant streaked photo and Auger electron emission is measured using a large-acceptance time-of-flight (TOF) spectrometer.

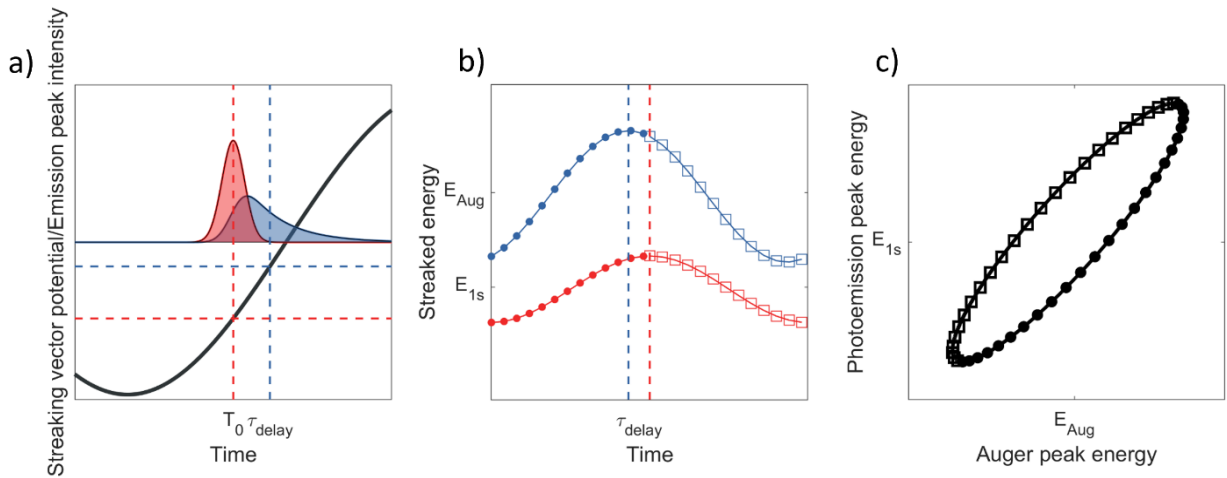

**Figure 2| Principle of self-referenced photoionised streaking measurements.** a) The photoelectrons (red) are emitted promptly after the arrival of the X-ray pulse at $T_0$. After the core-hole decays, the Auger electrons (blue) are emitted. The delay between the emission peaks' weighted centres, $\tau_{delay}$, which is highlighted by the distance between the vertical dotted lines, causes each set of electrons to interact with a different phase of the streaking pulse (solid black line). The horizontal dotted lines further highlight the difference in streaking field in each case. b) If the temporal overlap were smoothly varied over one streaking cycle, the resultant kinetic energy of each peak would trace out sinusoidal curves, shifted by the Auger decay lifetime. Filled circles represent positive streaking slopes at the moment of ionisation and open squares represent negative slopes. The dotted lines highlight the temporal overlap which results in the largest increase in kinetic energy for each peak, so that the gap between the dotted lines corresponds to the temporal shift between the two sinusoidal curves. c) Plotting each pair of positions against each other results in an ellipse. Filled circles and open squares have the same meaning as in (b).

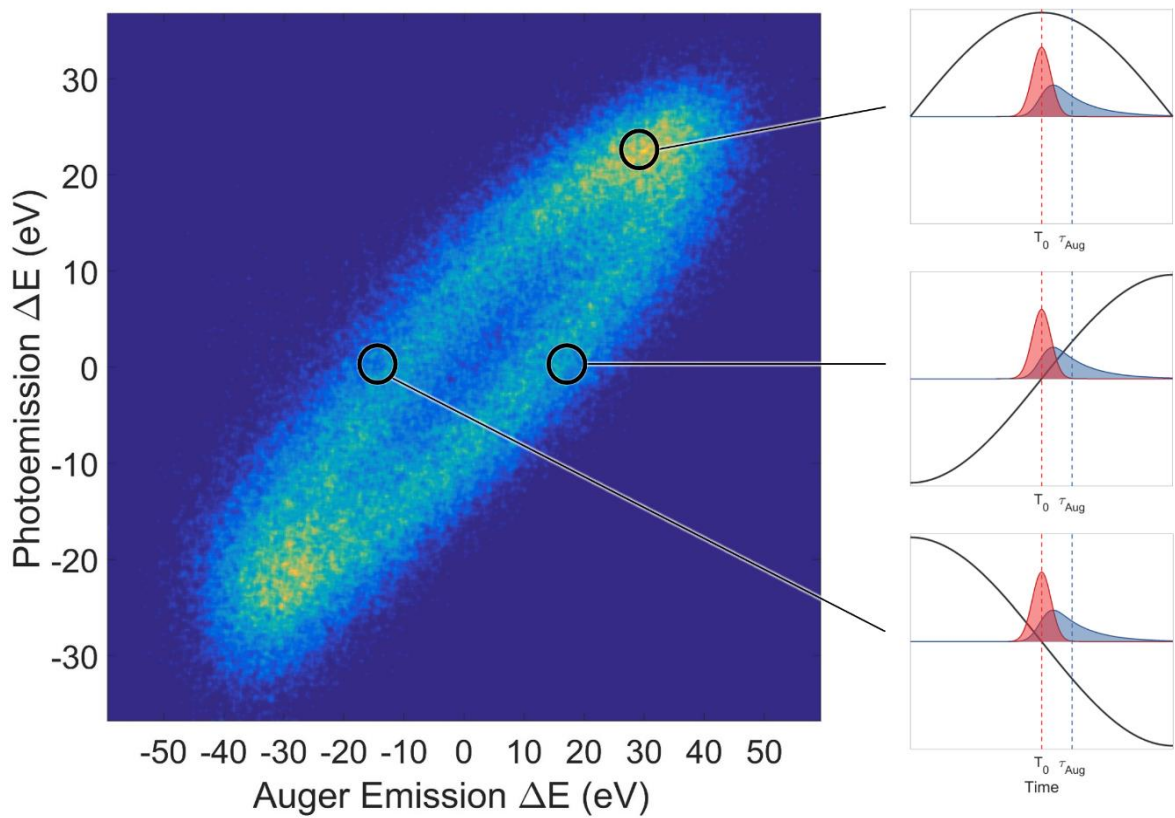

**Figure 3| Self-referenced streaking in neon.** Correlation map generated from 80,000 single-shot streaking measurements in neon using a 17μm streaking field, and 9.2 fs FWHM, 1130 eV ionising X-ray pulses. The *x* and *y* coordinates of the individual points in the scatter plot are determined by numerically fitting the streaked kinetic energy shift of the photo- and Auger electron peaks in each shot and calculating their centres of energy. On the right, three sketches are shown, corresponding to three characteristic regions on the map. The sketches show the photoelectrons (red) and Auger electrons (blue) along with the streaking field (black). The weighted centres of each temporal emission profile are highlighted with vertical dotted lines.

# Clocking Auger electrons: Supplementary information

**Theoretical background to the experimental method**

In order to verify the validity of our data analysis procedure, we performed a quantum-mechanical simulation of two-colour streaking for both photo- and Auger electrons. In this mathematical treatment of our experiment, we will consider the photoionisation of the neon 1s shell by a femtosecond, linearly polarised X-ray pulse. Photo- and Auger emission occur in the presence of a linearly polarised infrared (IR) field, synchronised with the ionising FEL pulse. It is assumed that the two beams are collinear and polarised along the z-direction.

The photo- and Auger electrons are not detected in coincidence, so they are not coherent and hence may be considered independently. Since both types of electrons are relatively fast ($E_{el} > 1$ a.u.), one can apply the Strong Field Approximation (SFA) [1, 2], wherein the probability of emission of a photoelectron with momentum $\vec{k}$ can be written as

$$W_{ph}(\vec{k}) = C \left| \int_{-\infty}^{\infty} dt \, \widetilde{\varepsilon_X(t)} \, D_{\vec{k}} \exp[i\Phi_{ph}(\vec{k},t)] \right|^2. \tag{8}$$

Here and in the following, all quantities are given in atomic units, unless otherwise stated. In equation (8), $\widetilde{\varepsilon_X(t)}$ is the envelope of the XUV pulse, $D_{\vec{k}}$ is the dipole matrix element describing the transition of the electron from the ground state to the continuum, $C$ is a constant which does not affect the following discussion, and $\Phi_{ph}(\vec{k},t)$ is related to the Volkov phase accumulated by the photoelectron as it moves in the IR field [3]. This can be written as

$$\Phi_{ph}(\vec{k},t) = -\int_{t}^{\infty} dt' \left[ \frac{1}{2}\left(\vec{k} - \overrightarrow{A_{IR}}(t')\right)^2 + (E_b - \omega_X) \right], \tag{9}$$

where $E_b$ is the absolute value of the photoelectron binding energy and $\omega_x$ is the carrier frequency of the XUV pulse, so that the energy of the photoelectron in the absence of the IR pulse $E_{ph}$ is given by $E_{ph} = \omega_X - E_b$. $\overrightarrow{A_{IR}}(t)$ is the vector potential of the IR laser field with electric field vector $\overrightarrow{\varepsilon_{IR}}(t) = \overrightarrow{\varepsilon_0} \cos(\omega_{IR} t)$, such that

$$\overrightarrow{A_{IR}}(t) = -\int_t^\infty dt' \overrightarrow{\varepsilon_{IR}}(t'). \tag{10}$$

Analogous to equation (8), we can calculate the probability of emission of an Auger electron of momentum $\vec{k}$ within the SFA [4]:

$$W_A(\vec{k}) = \frac{\Gamma}{8\pi} \left| \int_{t_0}^\infty dt \exp\left[i\Phi_A(t) - \frac{\Gamma(t-t_0)}{2}\right] \right.$$

$$\left. * \int_{t_0}^t dt' \, \widetilde{\varepsilon_X(t')} \, D_{\vec{k}} \exp\left[i\left(\left(E_e - \frac{i\Gamma}{2}\right)(t'-t_0) - (\omega_X - E_b)t'\right)\right] \right|^2. \tag{11}$$

Here, $t_0$ is the moment the XUV pulse starts to interact with the system, $E_e$ is the energy of the correlated photoelectron, and $\Gamma$ is the width of the Auger state, obtained from the Auger lifetime $\tau_A$ by the relation $\tau_A = \frac{1}{\Gamma}$. The quantity $\Phi_A(t)$ is defined as

$$\Phi_A(t) = -\int_t^\infty dt' \left[\frac{1}{2}\left(\vec{k} - \overrightarrow{A_{IR}}(t')\right)^2 - E_A\right], \tag{12}$$

where $E_A$ is the kinetic energy of the Auger electron in the absence of the IR field. Note that this is of a similar form to equation (9); in fact, both equations (8) and (11) can be written in the form

$$W_{el}(\vec{k}) = C \left| \int_{-\infty}^\infty dt \exp[i\Phi_{el}(\vec{k},t)] G_{el}(t) \right|^2, \tag{13}$$

with

$$\Phi_{el}(\vec{k},t) = -\int_t^\infty dt' \left[\frac{1}{2}\left(\vec{k} - \overrightarrow{A_{IR}}(t')\right)^2 - E_{el}\right]. \tag{14}$$

In equations (13) and (14), the subscript $el$ refers to the type of electrons being described – that is, either to photoelectrons $ph$ or Auger electrons $A$. For the photoelectrons, the factor $G_{el}(t)$ is given by

$$G_{ph}(t) = D_{\vec{k}}\widetilde{\varepsilon_X(t)}. \tag{15}$$

Henceforth, we shall set the dipole matrix element to unity and assume a simple Gaussian form for the X-ray pulse. Therefore,

$$G_{ph}(t) \approx \exp\left[-\frac{(t-\tau)^2}{2\sigma^2}\right], \tag{16}$$

where $\tau$ is the delay of the X-ray pulse with respect to the IR pulse, which varies stochastically from shot to shot. For the Auger case, the factor $G_A(t)$ is dependent on the autoionising Auger state and its decay linewidth $\Gamma$:

$$G_A(t) = \sqrt{\frac{\Gamma}{2\pi}} \exp\left(-\frac{\Gamma}{2}t\right) \int_{t_0}^{t} dt' \exp\left(\frac{\Gamma}{2}t'\right) \tilde{\varepsilon}_X(t'). \tag{17}$$

We will refer to the factors $G_{el}(t)$ as effective pulses. The square of the effective pulse is equal to the corresponding emission profile. The electrons' final momenta will depend upon the vector potential of the IR pulse, which is given by

$$A_{IR}(t) = A_0(t)\sin(\omega_{IR}t), \tag{18}$$

where $\omega_{IR} = \frac{2\pi}{T_{IR}}$ is the angular frequency and $T_{IR}$ is the period of the laser pulse. $A_0(t)$ represents the amplitude of the IR pulse. For simplicity, we will assume that the IR pulse is much longer than both the XFEL pulse and Auger decay lifetime. The result of this assumption is that for a *single shot*, the amplitude of the electric field interacting with photo- and Auger electrons can be assumed to be identical. However, $A_0(t)$ does vary on a timescale comparable to the timing jitter between X-ray and laser pulses, with the result that electrons emitted in *different* shots will generally interact with a different streaking amplitude. In Figure S1, the functions $G_{ph}^2$ and $G_A^2$ are shown, with all parameters set to values matching our experimental conditions.

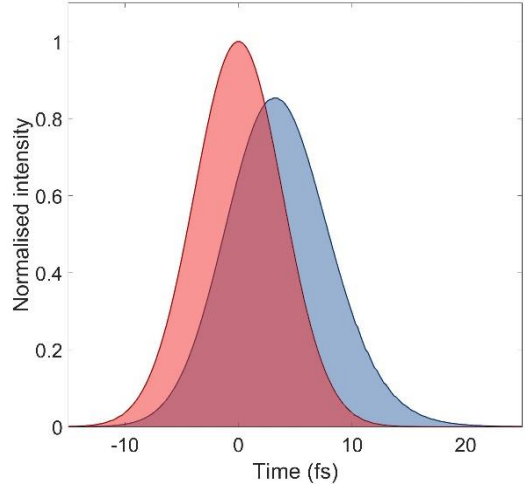

**S1: Simulated photo- and Auger emission profiles|** The red curve represents the photoemission profile, $G_{ph}^2(t)$, and the blue curve represents the Auger emission profile, $G_A^2(t)$. The time axis is relative to the time of arrival of the XFEL pulse. The delay between X-ray ionisation and the maximum of $G_A^2(t)$ is about 130 a.u. or 3.3 fs.

**The phenomenological approach**

It is worth comparing the SFA-based theory described above with the phenomenological '*ad hoc*' theory which has been applied in the past [5]. The latter is based on a description of the Auger process in terms of the following rate equation for the resonant state population:

$$\frac{dn(t)}{dt} = -\Gamma n(t) + \tilde{S}(t), \tag{19}$$

with solution

$$n(t) = \int_{-\infty}^{t} dt' \exp\left(-\Gamma(t - t')\tilde{S}(t')\right). \tag{20}$$

In equations (19) and (20), $n(t)$ represents the population of the Auger state, and $S(t)$ is a source term, usually defined as the probability of photoexcitation.

The crucial difference between the ad hoc and quantum approaches is that the dynamical quantity of the latter is the amplitude of the states, rather than their population. In the quantum model, the population of the state is computed by integrating the amplitude of the states over time and squaring the result. This difference in treatment means that the results predicted by the two models will generally differ. However, in the limit of very prompt excitation of the resonant state, both models predict an exponential decay in this state's population and will give similar results.

**Quasiclassical approach**

Whilst the final results presented in the main text were calculated using a fully quantum-mechanical approach, it is useful to first discuss a semiclassical approximation, which will be used to formulate a relationship between the final energy of an emitted electron and its moment of emission.

The phase $\Phi(\vec{k}, t)$ in equation (14) varies rapidly in time, making direct computation of the state amplitudes time-consuming. However, the fast oscillation of the integrand in equation (13), which describes the probability of electron emission, means the result will be dominated by the saddle points, where the phase is stationary in time. These points $t_s$ are defined by the condition $\frac{\partial \Phi(\vec{k},t)}{\partial t}|_{t=t_s} = 0$, with solutions given by

$$\left(\frac{k^2}{2} - E_{el}\right) - k \cos(\theta) A_{IR}(t_s) + \frac{A_{IR}^2(t_s)}{2} = 0, \qquad (21)$$

where $k = |\vec{k}|$, $A_{IR}(t) = |\vec{A_{IR}}(t)|$, and $\theta$ is the angle of electron emission. Henceforth, we shall assume that the electrons are detected along the direction of polarisation of the pulses, so that $\theta = 0$.

Equation (21) links the final momentum $k$, and hence final electron energy $E = \frac{k^2}{2}$, with the time of electron emission $t_s$. Solving it, we find

$$k - A_{IR}(t_s) = \sqrt{2E_{el}} = k_0; \qquad t_s = \frac{\arcsin\left[\frac{k - k_0}{A_0}\right]}{\omega_{IR}} + 2n\pi, \qquad n \in \mathbb{Z}, \tag{22}$$

where we have used the fact that for our experimental conditions, $A_0 \ll k, k_0$. In our experiment there is only one stationary point of $\Phi$ – the closest one to the excitation time $\tau$ – which contributes to the integrand in equation (13). Were the IR carrier frequency significantly higher, multiple stationary points would contribute, making the physical picture more complicated. This case is omitted from our discussion for brevity but could prove worthwhile for future investigation.

**Relationship between theoretical and experimentally measured quantities**

In the experiment, time-of-flight (TOF) spectra for photoelectrons and Auger electrons are recorded for every shot. Each measurement is made with a different $\tau$ and $A_{IR}(\tau)$, because neither the relative arrival time of the two pulses nor the carrier-envelope phase of the streaking pulse are controlled. This affects the shapes of the pair of TOF spectra obtained for each shot. After converting the spectra from TOF to kinetic energy, we evaluate the centre-of-energy (COE) for each of the two emission peaks. Plotting the change in COE, $\delta C_E$, for each of the two peaks against one another results in the elliptical figure shown in the main text. Measuring the phase shift between this ellipse's parametric components is how we arrive at the time-delay between the two centres, and we will use this value to calculate the Auger decay lifetime.

Within the full quantum treatment of the experiment, we can compute the change in an emission peak's centre of energy due to interaction with the streaking laser:

$$\delta C_E = \frac{\int \left(\frac{k^2}{2} - E_{el}\right) W_{el}(k) dk}{\int W_{el}(k) dk}. \tag{23}$$

Using the semiclassical approximation, we can transform the above expression and integrate over time instead of over emitted electron energy. Following the relations (22) and assuming that $A_{IR}(t) \ll k_0$, we find

$$\delta C_E = k_0 \frac{\int A_{IR}(t) G_{el}^2(t) dt}{\int G_{el}^2(t) dt}. \tag{24}$$

It is straightforward to evaluate this expression for the photoelectrons; the photoemission profile $G_{ph}(t)$ is short compared to the period of the streaking pulse, since $\sigma \ll T_{IR}$. The result is that the streaking vector potential varies slowly compared to the timescale of photoemission, allowing us to obtain a simple approximation for the photoelectron case:

$$\delta C_E \approx k_0 A_{IR}(\tau) = A_0 \sqrt{2E_{ph}} \sin(\omega_{IR}\tau). \tag{25}$$

The few-femtosecond Auger emission is also short compared to the period of the streaking pulse. Assuming a small variation in vector potential during Auger emission, one can expand $A_{IR}(t)$ about the XFEL arrival time $\tau$ using a Taylor series. Let the centre of time (COT) of the Auger emission profile be defined as

$$C_T = \frac{\int t G_A^2(t) dt}{\int G_A^2(t) dt}. \tag{26}$$

Note that this quantity is independent of $\tau$. We can use this to obtain an approximate formula for $\delta C_E$:

$$\delta C_E(A) \approx k_0 A_{IR}(\tau + C_T) = A_0 \sqrt{2E_A} \sin(\omega_{IR}\tau + \phi), \tag{27}$$

where $\phi = \omega_{IR} C_T$. This approximation provides a clear relation between the energy shift $\delta C_E$ and the COT $C_T$ of the effective Auger pulse. In this way, we can relate the spectral-domain quantities measured in the experiment to the temporal properties of the decay process.

**Calculation of the pulse duration**

The duration of the X-ray pulse is an important parameter in this experiment, as it influences the emission profile of both types of electrons. We will calculate it following the methods described in reference [6]. In streaking experiments such as ours, the duration of an X-ray pulse can be calculated using the relation

$$\tau_X = \frac{\sigma_{DC}}{s}, \quad (28)$$

where $\sigma_{DC}$ represents the breadth of the streaked photoemission peak after deconvolving that of the field-free peak, and $s = \frac{dE_k}{dt}$ is the streaking speed. The latter represents the rate of change of the peak's kinetic energy with respect to the timing of the streaking pulse. Where the X-ray and streaking pulses are well overlapped, the final energy of photoelectrons emitted at time $t$ is given by

$$E_k(t) = \frac{\Delta E_{max-min}}{2}\sin(\omega_{IR} t) + E_{1s}, \quad (29)$$

where $\Delta E_{max-min}$ is the difference between the most positive and most negative changes in photoelectron energy throughout the experiment, $\omega_{IR}$ is the angular frequency of the streaking field, and $E_{1s}$ is the field-free kinetic energy of the photoelectrons. Therefore, at the zero-crossing of the streaking field,

$$s = \frac{\Delta E_{max-min} * \pi}{T_{IR}}, \quad (30)$$

where we have used the relation $T_{IR} = \frac{2\pi}{\omega_{IR}}$ and set $t = 0$.

The value of $\Delta E_{max-min}$ is calculated from angular sectors corresponding to the maxima and minima of the streaking phase. All shots within a given angular sector interacted with the same streaking phase, but the final kinetic energy of the electrons will still vary according to the streaking field amplitude – which will itself vary due to timing jitter. Clearly, the largest possible change in kinetic energy will occur when both the streaking amplitude and phase are maximal. Because timing jitter results in a normal distribution of streaking amplitudes, it is most likely that a given shot will intersect with the centre of the streaking pulse envelope. These conditions are identified by numerically fitting the distribution of absolute photoelectron kinetic energies within both sectors and extracting its peak, as shown in Figure S2. Using this method, we determine that $\Delta E_{max-min} = 48 \pm 1$ eV.

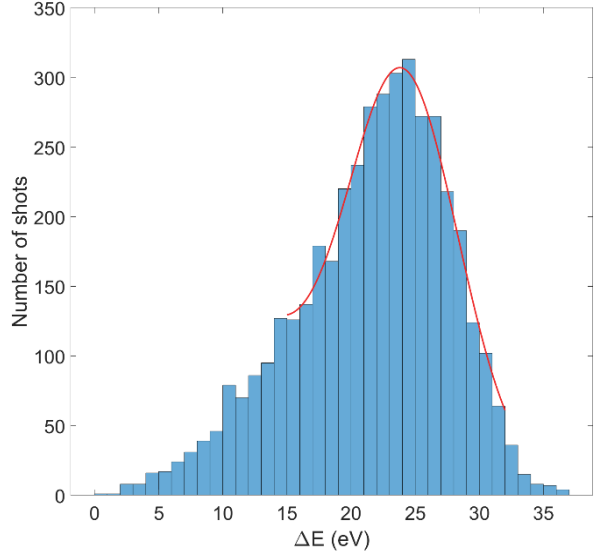

**S2: Distribution of maximally streaked kinetic energies|** The changes in photoelectron kinetic energy in the sector corresponding to maximal streaking phase are plotted in the histogram. The red line shows the numerically determined least-square fit, from which we extract the peak of the distribution. By the normally distributed nature of timing jitter, highest number of shots will overlap at or near the peak of the pulse envelope, so that the peak of the histogram ought to correspond to those conditions.

The first step towards finding $\sigma_{DC}$ is to calculate the spectral width of the photoemission peak at the zero-crossing of the streaking field. These shots can be swiftly identified using the elliptical distribution: they must lie on its equator, where the kinetic energy of the photoelectrons was largely unchanged. Further, the shots closest to the edge of the ellipse interacted with the peak of the streaking pulse envelope, resulting in a maximised change in Auger electron kinetic energy.

Therefore, we restrict our consideration to shots for which the final photoelectron kinetic energy was within 1 eV of its field-free value. Within this group, we take the 300 outermost shots on each side of the ellipse, corresponding to the strongest streaking effect. Recall that, as illustrated in Figure 3 of the main text, the left and right sides of the ellipse correspond to zero crossings of the streaking pulse with opposite slopes. The average breadth of the 300 shots on the left of the ellipse is $\sigma_L = 8.9 \pm 0.2$ eV and that of the 300 on the right is $\sigma_R = 9.8 \pm 0.2$ eV. Here and in the following, we have used the standard error on the mean value of $N$ repeated measurements, given by

$$\alpha_M = \frac{\delta}{\sqrt{N}}, \quad (31)$$

where $\delta$ is the statistical width of the distribution. We must deconvolve the bandwidth of the field-free photoemission peak $\sigma_{FF}$ from that of the peak at a zero-crossing using the relation

$$\sigma_{DC} = \sqrt{\sigma_S^2 - \sigma_{FF}^2} = 7.9 \pm 0.2 \text{ eV}. \quad (32)$$

Here, $\sigma_S^2 = \frac{\sigma_L^2 + \sigma_R^2}{2}$ represents the average of the squares of the mean bandwidth at each zero-crossing. The average field-free photoemission bandwidth is $\sigma_{FF} = 4.95 \pm 0.01$ eV. The uncertainty on $\sigma_{FF}$ is much smaller than that on $\sigma_S$, because there are many more unstreaked shots available to use in the calculation of the former. The XFEL pulse duration can be calculated using these quantities, propagating the uncertainties on each one using a standard functional approach. The full-width at half-maximum (FWHM) duration of the X-ray pulse is given by

$$\tau_X = 2\sqrt{2\ln(2)}\, \frac{\sigma_{DC}}{s} = 7 \pm 1 \text{ fs}, \quad (33)$$

where the factor $2\sqrt{2\ln(2)}$ is used to convert from rms width to FWHM.

**Extraction of the phase shift**

As noted in the main text, the two equations

$$x(\theta) = A\sin(\theta + \phi_A), \quad y(\theta) = B\sin(\theta) \quad (34)$$

describe an ellipse and plotting them against each other will allow us to measure $\phi_A$. A generalised plot of the two equations is shown in Figure S3. In this plot the $y$-intercept, $y_1$, as well as the maximum value of $y$, $y_2$, are highlighted.

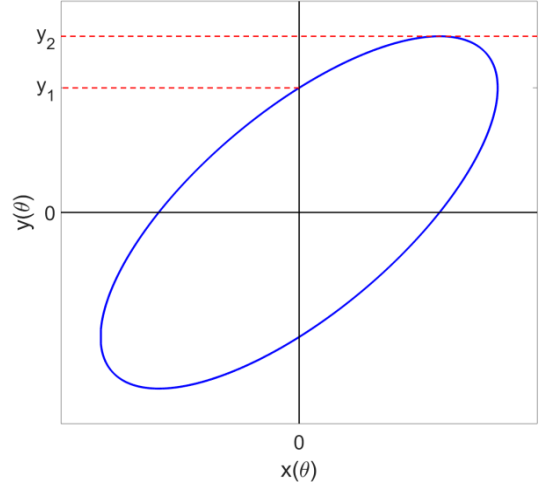

**S3: Generalised ellipse|** An arbitrary ellipse is shown (blue line) with the $x$- and $y$-axes highlighted (black lines), in addition to the parameters $y_1$ and $y_2$ (red dotted lines).

From equations (34), it is clear that

$$y_1 = B \sin(-\phi_A), \qquad y_2 = B. \tag{35}$$

It follows that

$$|\phi_A| = \sin^{-1}\left(\frac{y_1}{y_2}\right). \tag{36}$$

Therefore, we can measure the magnitude of $\phi_A$ simply by measuring the $y$-intercept and maximum value of $y$ in our data. In fact, the same principle can be applied to the negative $y$-intercept and extremum, and both positive and negative pairs on the $x$-axis. All four possible measurements were made and showed little disagreement. For the result shown in the paper, the positive $y$-pair was used. The motivation for choosing $y_1$ and $y_2$ as opposed to $x_1$ and $x_2$ comes from the fact that our spectral resolution on the photoemission ($y$) peak is better than that of the Auger peak. Furthermore, $x_1$ ought to be measured at a zero crossing of the photoemission peak. Under these conditions the peak will be significantly broadened, and its position will be more uncertain. In contrast, the Auger peak – and therefore measurements using points on the $y$-axis, which is at a zero crossing of the Auger peak – is less sensitive to broadening-induced noise.

**Error analysis**

The method lends itself to straightforward error analysis. If we are able to quantify the uncertainties on $y_1$ and $y_2$ as $\alpha_1$ and $\alpha_2$ respectively, we could define a parameter $K = \frac{y_1}{y_2}$, whose uncertainty will be given by the following expression:

$$\alpha_K = K\sqrt{\left(\frac{\alpha_{y_1}}{y_1}\right)^2 + \left(\frac{\alpha_{y_2}}{y_2}\right)^2}. \tag{37}$$

From here it is simple to quantify the uncertainty on $\phi_A$ using a functional approach and equation (36), as follows:

$$\alpha_\phi^\pm = |\sin^{-1}(K) - \sin^{-1}(K \pm \alpha_K)|. \tag{38}$$

Recall that $\phi_A$ corresponds to the streaking phase advance which occurs between the centres of energy of the photo- and Auger emission bursts. As described in the main text, the centre-of-mass delay between the two emissions, $\tau_{delay}$, is a scalar product of $\phi_A$ and the MIR streaking laser wavelength $\lambda_{IR}$:

$$\tau_{delay} = \frac{\phi_A \lambda_{IR}}{2\pi c}. \tag{39}$$

The constant $2\pi c$ can be assumed to be known to an infinite degree of precision. It is now straightforward to calculate $\alpha_{delay}^\pm$, the uncertainty on $\tau_{delay}$, which is given by

$$\alpha_{Aug}^\pm = \tau_{delay}\sqrt{\left(\frac{\alpha_\phi^\pm}{\phi_A}\right)^2 + \left(\frac{\alpha_\lambda^\pm}{\lambda}\right)^2}. \tag{40}$$

The wavelength $\lambda$ was measured at LCLS to be $17 {}^{+1}_{-2}$ μm. Therefore, we need only identify $\alpha_1$ and $\alpha_2$ in order to determine the precision of our measurement of the centre-of-mass delay. The quantities $y_1$ and $y_2$, and their uncertainties, were measured by examining the distribution of data in particular angular sectors of the ellipse.

**Selection of ellipse sectors**

It is clear that, when measuring from the origin, the ellipse's $y$-intercept $y_1$ is contained within a sector aligned with the positive $y$-axis. It is less obvious to determine which sector contains $y_2$, the maximum value of $y$.

From the general equation of an ellipse,

$$\begin{pmatrix} x(t) \\ y(t) \end{pmatrix} = \begin{pmatrix} \cos(\alpha) & -\sin(\alpha) \\ \sin(\alpha) & \cos(\alpha) \end{pmatrix} \begin{pmatrix} a\cos(t) \\ b\sin(t) \end{pmatrix} + \begin{pmatrix} x_0 \\ y_0 \end{pmatrix}, \tag{41}$$

one can derive the angle $\theta_c$ for which $y$ is maximal:

$$\theta_c = \tan^{-1}\left(\frac{b}{a}\cot(\alpha)\right). \tag{42}$$

It follows that $y_2 = y(\theta_c)$. An initial approximation to $\theta_c$ was made using a least-squares fit of the elliptical data, which provided the necessary constants $\alpha, a$ and $b$ and resulted in a value of $\theta'_c = 0.6612$ rad. To support this choice of $\theta_c$, $y_2$ was measured for a range of critical angles $\theta_c$. The results are visible in Figure S4.

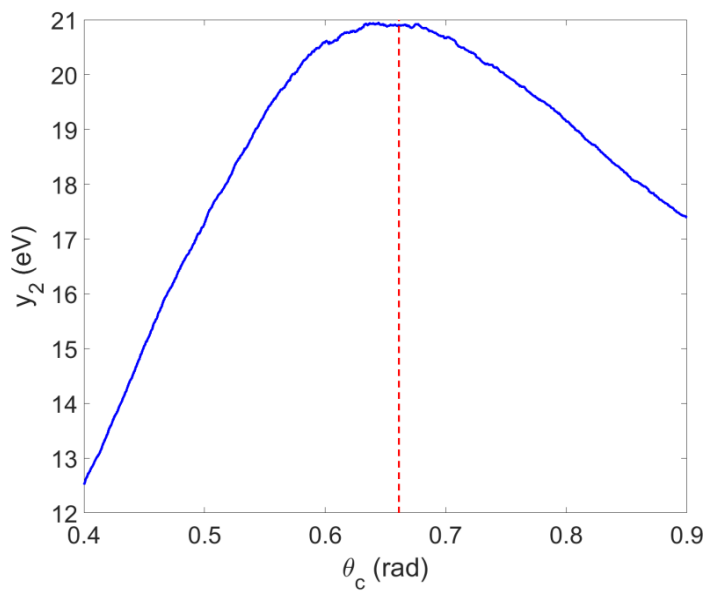

**S4: Critical angle $\theta_c$** | The measured value of $y_2$ is shown (blue line) for a range of critical angles $\theta_c$. The red dotted line represents the value of $\theta_c$ which was used in the final analysis.

The value of $\theta_c$ which maximises $y_2$ was found to be very close to that determined from the fit. Furthermore, $y_2$ shows little variation in the region of $\theta_c$-space close to $\theta_c'$. One can infer that this was a good choice of $\theta_c$, and further that any uncertainty on $\theta_c'$ will have a small impact, because it will result in only a small change to $y_2$. The next step is to identify $y_2$ from the distribution of points in the sector at $\theta_c$. The two sectors are highlighted in Figure S5.

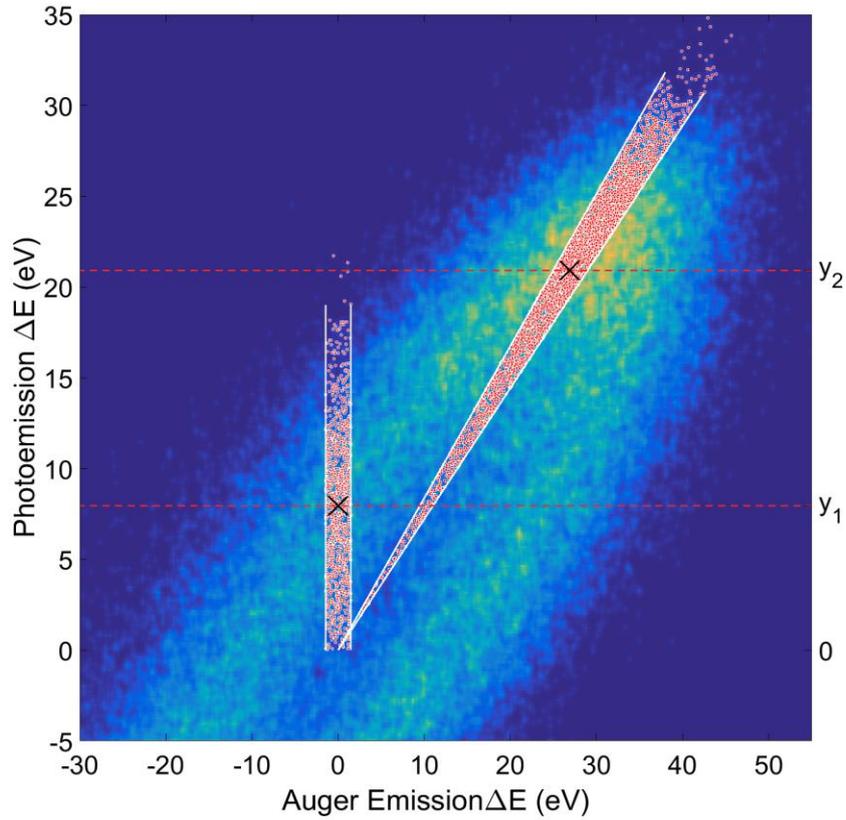

**S5: Sectors for calculation of Lissajous parameters $y_1$ and $y_2$.** A zoomed-in section of the data is shown. Overlaid on the density map, the red points are those contained in each of the sectors used to find the Lissajous parameters. The black crosses represent the measured values $y_1$ and $y_2$, which are highlighted by the red dotted lines. From these data, $y_1$ and $y_2$ were measured at 7.9 and 20.9 eV respectively.

Let the sector from which we obtain the value $y_i$ be called $Y_i$, such that $y_i$ is given by the mean of all the $y$-coordinates of the points in $Y_i$. The sectors $Y_1$ and $Y_2$ are defined in a subtly different way. By definition, $y_1$ is the $y$-intercept of the ellipse, and as such the sector $Y_1$ is identified by taking the set of points closest to the $y$-axis, so that

$$Y_1 = \{\, (x, y) : |x| < \varepsilon_1 \,\}, \qquad \varepsilon_1 \in \mathbb{R}, \varepsilon_1 > 0. \tag{43}$$

The sector $Y_2$, conversely, is defined as the points whose angular coordinate is closest to the critical angle $\theta_c$. In terms of polar coordinates, we can define $Y_2$ as follows:

$$Y_2 = \{ (r, \theta) : |\theta - \theta_c| < \varepsilon_2 \}, \qquad \varepsilon_2 \in \mathbb{R}, \varepsilon_2 > 0. \tag{44}$$

A polar sector is ill-suited for determining $y_1$, as it incorporates outliers at the edge of the cone, causing the algorithm to overestimate $y_1$ and therefore $\phi_A$ when tested with simulated data. However, $\theta_c$ is sufficiently large that the boundaries of $Y_2$ are almost perpendicular to the edge of the distribution, ensuring that few outliers are included. Through repeated tests with simulations, it was verified that a sector of this type results in the most accurate determination of $y_2$. This sector was also used in the calculation of the pulse duration, as highlighted in Figure S2.

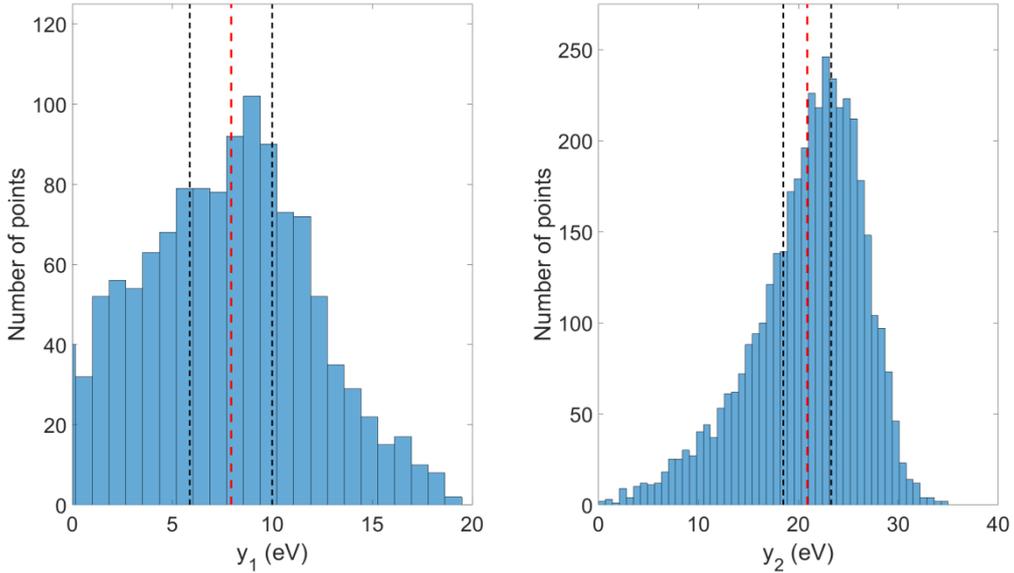

**S6: The distribution of parameters $y_1$ and $y_2$|** The red dashed lines highlight the mean value, while the black dashed lines display the statistical width of the distribution.

The distributions of $y_1$ and $y_2$ obtained from the sectors $Y_1$ and $Y_2$ are shown in Figure S6. One can consider the spread of $y$-values inside a sector to be in effect repeated measurements of $y_1$ and $y_2$. Therefore, taking the mean of all points in the sector gives a notion of the true values.

The sector $Y_1$ contains 1000 points and $Y_2$ contains 4000 points. As the number of points increases, the standard error given by equation (31) can be reduced. However, as we make the sectors wider and wider, the width of the distribution, $\delta$, begins to increase. The sector sizes were chosen to minimise $\alpha_M$, resulting in uncertainties of around 0.1eV for $y_1$ and $y_2$. The procedure described above can now be used to go from these uncertainties to those of the COE delay $\tau_{delay}$. The dominant source of uncertainty in the experiment is in the measurement of the streaking wavelength, which was only known within a 3 µm range. Accounting for this using equation (40) results in the value of $3.5^{+0.3}_{-0.5}$ fs given in the main text. This value can be interpreted in the context of the quantum theory above; we can simulate the experiment for a range of Auger decay lifetimes and calculate the resulting centre-of-mass delay $\tau_{delay}$. A functional approach allows us to propagate the uncertainties and arrive at our final result of $2.2^{+0.2}_{-0.3}$ fs for the Auger decay lifetime.

**References:**

1. Keldysh L.V. Ionization in the field of a strong electromagnetic wave. *Soviet Physics JETP* **20**, *1307 (1965)*.
2. Kazansky, A. K., Sazhina, I. P., and Kabachnik, N. M. Angle-resolved electron spectra in short-pulse two-color XUV+IR photoionization of atoms. *Phys. Rev. A* **82**, *033420 (2010)*.
3. Wolkow, D. M. Über eine Klasse von Lösungen der Diracschen Gleichung *Zeitschrift für Physik* **94**, *250 (1935)*.
4. Kazansky, A. K., Sazhina, I. P., and Kabachnik N. M. Time-dependent theory of Auger decay induced by ultra-short pulses in a strong laser field. *J. Phys. B* **42**, *245601 (2009)*.
5. Drescher, M. *et al.* Time-resolved atomic inner-shell spectroscopy, *Nature* **419** *803 (2002)*.
6. Itatini, J. *et al.* Attosecond Streak Camera. *Phys. Rev. Lett.* **88**, *173903 (2002)*.